\pdfoutput=1

\documentclass[11pt]{article}

\usepackage[utf8]{inputenc}
\usepackage[T1]{fontenc}
\usepackage{lmodern}
\usepackage[letterpaper,margin=1in]{geometry} 
\usepackage{microtype}
\usepackage{amsmath,amssymb}
\usepackage{graphicx}
\usepackage{booktabs,tabularx,array}
\usepackage{xcolor}
\usepackage{soul} 
\usepackage[font=small,labelfont=bf]{caption}
\usepackage[numbers,square,sort&compress]{natbib} 

\definecolor{darkred}{rgb}{0.9,0,0}
\usepackage[colorlinks=true,linkcolor=darkred,citecolor=darkred,urlcolor=darkred,breaklinks=true]{hyperref}\usepackage{doi} 
\usepackage{orcidlink} 

\newif\ifcompactinitials
\compactinitialstrue
\newcommand{\initials}[2]{\ifcompactinitials#1\else#2\fi}

\newcommand{\code}[1]{\texttt{#1}}
\newcommand{\package}[1]{\texorpdfstring{\textbf{\texttt{#1}}}{#1}}
\newcommand{\qem}{\package{quantEM}}
\definecolor{classcolor}{rgb}{0.5,0,1.0}
\newcommand{\class}[1]{\texorpdfstring{\textcolor{classcolor}{\texttt{#1}}}{#1}}

\renewenvironment{abstract}{\begin{center}\textbf{Abstract}\end{center}\vspace{-0.6em}}{}

\title{\qem{}: Quantitative Electron Microscopy Analysis from Differentiable Physics to Neural Priors}

\author{Colin Ophus\textsuperscript{1,$*$}\,\orcidlink{0000-0003-2348-8558}, Arthur \initials{RC}{R.~C.} McCray\textsuperscript{1,$\dagger$}\,\orcidlink{0000-0001-6077-4698},\\ Georgios Varnavides\textsuperscript{2,$\ddagger$}\,\orcidlink{0000-0001-8338-3323}, and Stephanie \initials{M}{M.} Ribet\textsuperscript{3,\S}\,\orcidlink{0000-0002-7117-066X}}
\date{}

\begin{document}

\maketitle

\vspace{-2.5em}
\begin{flushleft}
\small
\textsuperscript{1}Department of Materials Science and Engineering, Stanford University, Stanford, CA, USA\\
\textsuperscript{2}Department of Imaging Physics, Faculty of Applied Sciences, Delft University of Technology, Delft, The Netherlands\\
\textsuperscript{3}National Center for Electron Microscopy, Molecular Foundry, Lawrence Berkeley National Laboratory, Berkeley, CA, USA\\[4pt]
\textsuperscript{$*$}\href{mailto:cophus@stanford.edu}{cophus@stanford.edu}, \textsuperscript{$\dagger$}\href{mailto:amccray@stanford.edu}{amccray@stanford.edu}, \textsuperscript{$\ddagger$}\href{mailto:g.varnavides@tudelft.nl}{g.varnavides@tudelft.nl}, \textsuperscript{\S}\href{mailto:sribet@lbl.gov}{sribet@lbl.gov}\\[4pt]

\end{flushleft}

\begin{abstract}
Quantitative electron microscopy includes imaging, diffraction, spectroscopy, and tomography measurements, and the analysis of these measurements increasingly requires solving inverse problems with physical forward models.
Open-source software has made many of these analyses widely available, but most packages are organized around a single modality, so that multimodal analyses require moving data between packages, and new methods cannot easily reuse the physical models and optimizers of existing ones.
Here we present \qem{}, an open-source Python package for quantitative electron microscopy that uses \package{PyTorch} tensors for computation and automatic differentiation, and \package{Zarr} for storage.
We implement each analysis method by writing a differentiable forward model and choosing a representation, loss function, constraints, and optimizer from a library of components, so that every method runs on CPUs and GPUs, can be saved and resumed, and uses the same calibrated data structures and visualization tools.
We describe the current modules, including orientation and strain mapping, direct and iterative ptychography, tomography with neural representations, and X-ray and electron energy-loss spectroscopy, and show how we assemble combined methods such as ptychographic tomography from existing components.
\qem{} is available under the MIT license, together with documentation, tutorials, and curated example datasets.
\end{abstract}

\textbf{Keywords:} transmission electron microscopy, open-source software, automatic differentiation, inverse problems, multimodal analysis, GPU computing

\section{Introduction}
\label{sec:introduction}

\subsection{Quantitative Electron Microscopy}

Transmission electron microscopy (TEM) and scanning TEM (STEM) can measure the atomic structure, composition, bonding, and electromagnetic fields of materials over length scales from micrometers to atomic resolution~\cite{williams2009transmission,pennycook2011scanning}.
Modern TEM and STEM instruments support a variety of characterization modalities, including imaging, diffraction, spectroscopy, and 3D tomography~\cite{ophus2023quantitative}.
These methods can be further expanded to include time-series measurements during a wide variety of applied sample stimuli, including heating, cooling, electrical biasing, and straining~\cite{fan2019insitu}.

The volume of data collected has also dramatically expanded with the widespread adoption of high-speed direct electron detectors.
For example, \emph{in situ} or \emph{operando} experiments can record continuous movies over very large fields of view~\cite{mckeowngreen2026insitu}, and current detectors running at their maximum frame rates produce a terabyte of data in several minutes~\cite{datta2021data}.
Four-dimensional STEM (4DSTEM) experiments, where we record a 2D image of the diffracted probe over a grid of 2D positions~\cite{ophus2019four}, have massive data rates.
Spectroscopy data have also expanded in size, especially when we perform simultaneous multimodal imaging experiments and extend these measurements to tilt series to recover 3D chemistry~\cite{collins2017progress}.
These datasets can easily require hundreds of gigabytes of storage per experiment, and with modern detectors we may collect hundreds of these large datasets per day~\cite{ercius2024fourd}.
Quantitative analysis of these large TEM and STEM datasets requires robust analysis software, careful calibration, and in many cases solving nonconvex, underdetermined inverse problems~\cite{fienup1982phase,midgley2009electron}.
Automated acquisition and online analysis place additional demands on the speed, scalability, and reproducibility of analysis software~\cite{spurgeon2021towards}.

\subsection{Open-Source Software for Electron Microscopy}

Open-source software is widely used to analyze large-scale TEM and STEM datasets, especially given the widespread deployment of machine learning for data analysis~\cite{kalinin2022machine}.
Python in particular has become the \emph{de facto} programming language for both open-source scientific analysis~\cite{oliphant2007python,millman2011python} and machine learning~\cite{raschka2020machine,paszke2019pytorch}, including for TEM.
One of the first wide-ranging TEM data analysis toolkits was \package{HyperSpy}, a general Python framework for multidimensional electron microscopy signals with calibrated axes, lazy loading of large datasets, and model fitting, initially for electron energy-loss spectroscopy (EELS) and X-ray energy-dispersive spectroscopy (XEDS) spectrum images~\cite{delapena2017electron,hyperspy}.
The \package{HyperSpy} ecosystem now includes specialized packages such as \package{RosettaSciIO} for file input and output~\cite{rosettasciio}, \package{eXSpy} for spectroscopy~\cite{exspy}, \package{pyxem} for 4DSTEM~\cite{cautaerts2022free}, \package{orix} for crystal orientations~\cite{johnstone2020density}, and \package{Atomap} for atomic-resolution images~\cite{nord2017atomap}.
Another widely used tool for 4DSTEM analysis is \package{py4DSTEM}~\cite{savitzky2021py4dstem}, which includes virtual imaging, Bragg disk detection, strain mapping, orientation mapping~\cite{ophus2022automated}, and phase retrieval~\cite{varnavides2023iterative}, together with an HDF5-based file format.
We previously described the architecture, development cycle, and governance of \package{py4DSTEM} as an example of research software co-created by many contributors~\cite{savitzky2023architecture}, and many of the design decisions in \qem{} follow directly from that experience.
Similar lessons have been described for \package{pymatgen}~\cite{ong2013pymatgen}, one of the largest open-source codebases in materials science, after 15 years of community-driven development~\cite{george2026pymatgen}.
A third widely used TEM analysis package is \package{LiberTEM}, which processes large datasets with user-defined functions in a map-reduce framework that scales from a workstation to a cluster~\cite{clausen2020libertem}, and was later extended to live processing during 4DSTEM acquisition~\cite{strauch2021live,clausen2023live}.
Other open-source tools for TEM analysis include \package{pycroscopy} and USID~\cite{somnath2019usid}, \package{AtomAI} for deep learning~\cite{ziatdinov2022atomai}, \package{StatSTEM} (MATLAB) for model-based quantification of atomic-resolution images~\cite{debacker2016statstem}, and a package for real-time phase contrast imaging from sparse electron event data recorded by the 4D Camera~\cite{pelz2022realtime}.

Open-source 3D tomographic reconstruction codes that are not specific to TEM include \package{TomoPy}~\cite{gursoy2014tomopy}, the \package{ASTRA} toolbox~\cite{vanaarle2015astra,vanaarle2016fast}, \package{TIGRE}~\cite{biguri2016tigre}, and the Core Imaging Library (\package{CIL})~\cite{jorgensen2021cil}.
Electron tomography codes include \package{IMOD}~\cite{kremer1996imod}, \package{Tomo3D}~\cite{agulleiro2011tomo3d}, and \package{AreTomo}~\cite{zheng2022aretomo}, which are widely used for cryo-electron tomography, and \package{tomviz}~\cite{schwartz2022realtime}, \package{GENFIRE}~\cite{pryor2017genfire}, \package{RESIRE}~\cite{pham2023resire}, and \package{ETSpy}~\cite{herzing2025etspy} for materials science.
Open-source ptychography codes include \package{PtychoShelves}~\cite{wakonig2020ptychoshelves} and its electron ptychography extension \package{fold\_slice}~\cite{chen2021electron}, \package{PtyPy}~\cite{enders2016ptypy}, \package{PyNX}~\cite{favrenicolin2020pynx}, \package{PtyLab}~\cite{loetgering2023ptylab}, \package{PtyRAD}~\cite{lee2025ptyrad}, \package{Pty-Chi}~\cite{du2025ptychi}, \package{phaser}~\cite{gilgenbach2026phaser}, and \package{PtyRANNOSAUR}~\cite{loehr2026ptyrannosaur}.

Several open-source packages are widely used for TEM and STEM simulations.
These include \package{$\mu$STEM}~\cite{allen2015modelling}, \package{Dr.~Probe}~\cite{barthel2018drprobe}, \package{computem}~\cite{kirkland2020advanced}, \package{Prismatic}~\cite{ophus2017fast,rangeldacosta2021prismatic,pelz2021faster}, and \package{abTEM}~\cite{madsen2021abtem}, which is now the most widely used.
\package{abTEM} has been widely adopted because of its easily expandable and modular all-Python implementation, fast computational models, ease of use, implementation of many algorithms and electron scattering modalities, and integration with first-principles simulations in \package{GPAW}~\cite{mortensen2024gpaw,susi2019efficient}.

Nearly all of these packages are built on the scientific Python stack, including \package{NumPy}~\cite{harris2020array}, \package{SciPy}~\cite{virtanen2020scipy}, \package{Matplotlib}~\cite{hunter2007matplotlib}, \package{scikit-image}~\cite{vanderwalt2014scikit}, and \package{Jupyter} notebooks~\cite{kluyver2016jupyter}.
Many follow the FAIR principles for data and research software and the software citation principles~\cite{wilkinson2016fair,barker2022introducing,smith2016software}.
Most of these packages target one modality or one class of algorithms, and many implement the update rule of each reconstruction algorithm by hand, although recent ptychography codes use automatic differentiation~\cite{lee2025ptyrad,du2025ptychi,gilgenbach2026phaser}.
As a result, multimodal analyses require conversion between packages, and each new forward model often requires its gradients to be derived and implemented by hand.

\subsection{Differentiable Programming for Inverse Problems}

Automatic differentiation (AD) computes exact gradients of a numerical program by applying the chain rule to each elementary operation, at a computational cost comparable to evaluating the program itself~\cite{baydin2018automatic}.
AD therefore removes the need to derive the gradients of each new forward model by hand, and greatly simplifies phase retrieval algorithms for complicated forward models~\cite{jurling2014applications}.
AD has been applied to X-ray ptychography~\cite{nashed2017distributed,ghosh2018adp,kandel2019using}, and generalized to other X-ray modalities in \package{Adorym}~\cite{du2021adorym}.
In electron ptychography, AD has been used to reconstruct datasets with reduced information content using regularization~\cite{schloz2020overcoming} and to invert partially coherent dynamical scattering~\cite{diederichs2024exact}.
AD also underlies the \package{PyTorch}-based packages \package{PtyRAD} and \package{Pty-Chi}~\cite{lee2025ptyrad,du2025ptychi}.

Deep learning frameworks such as \package{PyTorch} provide AD, GPU acceleration, and distributed training~\cite{paszke2019pytorch,ansel2024pytorch,li2020pytorch}, and make it straightforward to represent an unknown object with a neural network.
Many different neural network architectures have been proposed, including deep image priors~\cite{ulyanov2018deep}, neural radiance fields~\cite{mildenhall2020nerf}, implicit neural representations (INRs) with periodic activations or Fourier features~\cite{sitzmann2020implicit,tancik2020fourier}, hash encoding~\cite{muller2022instant}, tensor decompositions such as K-Planes~\cite{fridovichkeil2023kplanes}, and many others.
Neural representations have reconstructed heterogeneous structures in cryo-EM~\cite{zhong2021cryodrgn} and noisy STEM tomography data~\cite{kniesel2022clean}.

We have applied these ideas to ptychography with deep generative priors~\cite{mccray2025deep,mccray2025accelerating}, INR tomography with joint alignment~\cite{lim2025missing}, three-dimensional multislice ptychography~\cite{ribet2024uncovering,varnavides2023iterative}, cryo-electron ptychography of proteins~\cite{kucukoglu2024low}, and phase-contrast cryo-tomography~\cite{lee2025phaset3m}, among other applications.

\subsection{Overview of This Work}

\begin{figure}[tbp]
\centering
\includegraphics[width=6.4in]{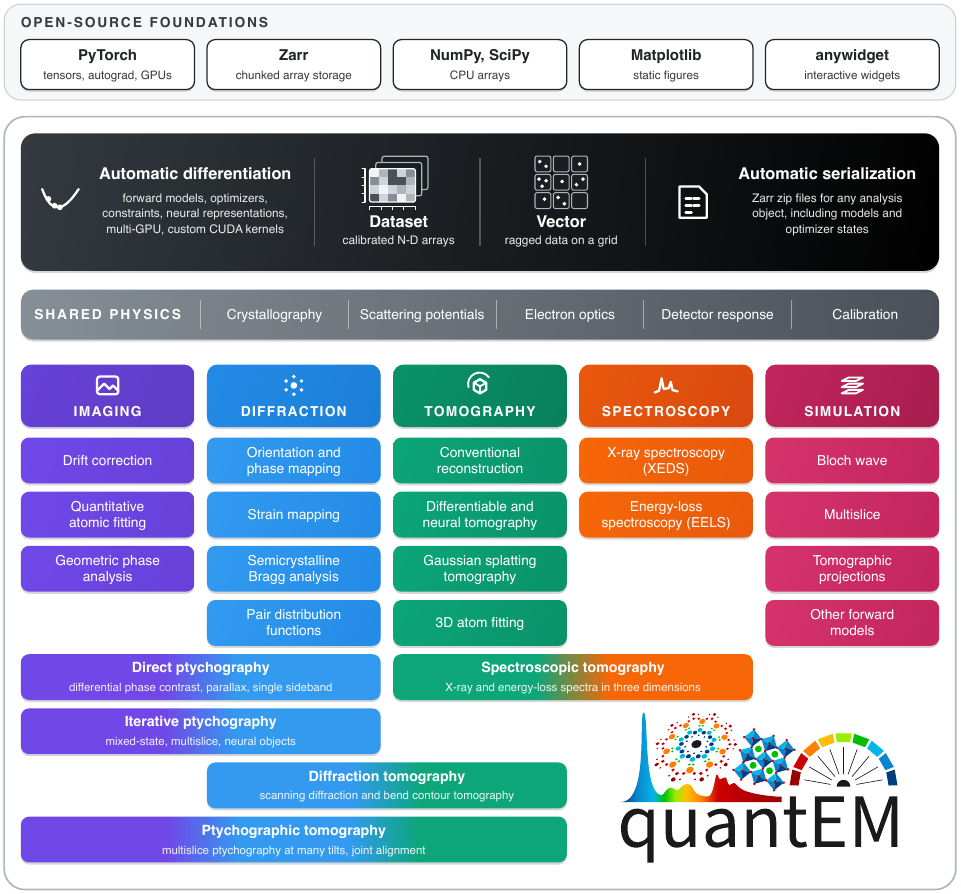}
\caption{Structure of \qem{}: open-source foundations, the \qem{} core, the shared physics layer, five domains, the analysis modules in each domain, and combined modules.}
\label{fig:architecture}
\end{figure}

In this paper, we describe \qem{}, an open-source Python package for quantitative electron microscopy built on differentiable physical models~\cite{quantem}, shown schematically in Figure~\ref{fig:architecture}.
We focus on its current modules and on how we implement new analysis methods.
Our goal is to make quantitative analysis of electron microscopy data calibrated, reproducible, and differentiable, and to reduce the effort needed to turn a new physical model into a working analysis method.
We write each analysis method in \qem{} as a forward model in \package{PyTorch}, so that the gradients of the loss with respect to all model parameters are computed by automatic differentiation rather than derived by hand, and the same code runs on CPUs, GPUs, and multiple GPUs.
Every dataset stores the origin, sampling, and units of each axis, every operation updates these calibrations, and we can save any analysis object, including neural network weights and optimizer states, to a single \package{Zarr} file, which we can share with others or reload later to resume the analysis.
We assemble new methods from shared components, including calibrated data structures, physics models, object representations, loss functions, constraints, and optimizers, so that we can build multimodal methods such as ptychographic tomography directly from the methods they combine.
We organize analysis methods into five domains (imaging, diffraction, tomography, spectroscopy, and simulation) connected by a shared physics layer.
We keep all modalities in a single repository under the permissive MIT license, and we intend \qem{} to be used alongside the existing open-source tools described above.

Section~\ref{sec:structure} describes the overall structure of the project and the design decisions behind it (Sections~\ref{sec:foundations}--\ref{sec:design}), the \class{Dataset} and \class{Vector} classes (Section~\ref{sec:core}), differentiable forward models (Section~\ref{sec:forward}), serialization (Section~\ref{sec:serialization}), visualization (Section~\ref{sec:visualization}), and the building blocks for new analysis methods (Section~\ref{sec:new}).
Section~\ref{sec:modules} describes the current modules in each domain.
Section~\ref{sec:discussion} compares \qem{} to other open-source packages (Section~\ref{sec:comparison}), describes how we develop it (Section~\ref{sec:contributing}), and gives the development status and roadmap (Section~\ref{sec:status}).
Finally, Section~\ref{sec:conclusions} summarizes this work.
We describe several \qem{} modules in more detail in separate method papers, including direct ptychography~\cite{varnavides2026relaxing} and multi-angle precession electron diffraction~\cite{ribet2025maped}.
Here we summarize each module with references to the method papers.

\section{Project Structure}
\label{sec:structure}

Figure~\ref{fig:architecture} shows the layers of \qem{} from the most general to the most specialized: open-source foundations, the core class functionality, a shared physics layer, five domains, the modules within each domain, and combined modules.
The figure shows the planned scope, and Table~\ref{tab:status} gives the current status of each module.

\subsection{Open-Source Foundations}
\label{sec:foundations}

We use \package{PyTorch} to provide the tensors used by every module, on CPUs, NVIDIA GPUs (CUDA), and Apple silicon GPUs (MPS), together with reverse-mode AD, neural network layers, optimizers, learning-rate schedulers, and communication between GPUs and nodes~\cite{paszke2019pytorch,ansel2024pytorch,li2020pytorch}.
We use \package{Zarr} (version 3) to store N-dimensional arrays as compressed chunks in a directory or a single zip file~\cite{zarr}.
Supporting libraries include \package{NumPy} and \package{SciPy}~\cite{harris2020array,virtanen2020scipy}, \package{Matplotlib}~\cite{hunter2007matplotlib}, \package{scikit-image}~\cite{vanderwalt2014scikit}, \package{RosettaSciIO} for file input and output~\cite{rosettasciio}, \package{Optuna} for hyperparameter optimization~\cite{akiba2019optuna}, and \package{anywidget} for interactive widgets~\cite{manz2024anywidget}.

\subsection{Core, Physics Layer, and Domains}

The core provides automatic serialization to every module, and defines two Python classes for data: \class{Dataset} for calibrated N-dimensional arrays and \class{Vector} for ragged data on a fixed grid (Section~\ref{sec:core}).
Many modules implement automatic differentiation, and a shared physics layer collects the models needed by several domains: crystallography, electron scattering factors and potentials, electron optics, detector response, and calibration.
Most modules fall into imaging, diffraction, tomography, and spectroscopy categories~\cite{ophus2023quantitative}.
Simulation provides forward models for the other domains, and we build combined modules such as diffractive imaging, ptychographic tomography, and XEDS and EELS tomography from the classes of the domains they combine.
We package each module under a descriptive name, such as \package{quantem.imaging}, \package{quantem.diffraction}, \package{quantem.diffractive\_imaging}, \package{quantem.tomography}, and \package{quantem.spectroscopy}.

\subsection{Design Decisions}
\label{sec:design}

We use \package{PyTorch} as the single compute backend.
Using a single backend avoids the interchangeable \package{NumPy} and \package{CuPy} code paths of \package{py4DSTEM}~\cite{savitzky2021py4dstem}, which in our experience increased the testing effort and made GPU support uneven across modules.
We chose \package{PyTorch} over \package{JAX}, which offers similar AD and accelerator support with a functional programming model, because its object-oriented modules map onto the object, probe, and detector components of a forward model, it supports Apple silicon GPUs, and it is widely used in microscopy and machine learning~\cite{ziatdinov2022atomai,zhong2021cryodrgn,lee2025ptyrad,du2025ptychi}.
The costs are a large dependency and a higher per-operation overhead than \package{NumPy} for small arrays.
Our goal is to run every module on \package{PyTorch}, and we are converting the few modules that still use \package{NumPy} and \package{SciPy}, such as drift correction and lattice fitting.
The chunked \package{Zarr} format with JSON metadata allows partial reads of large arrays, parallel writes, and cloud object storage~\cite{zarr}, and the bioimaging community has adopted it as the basis of OME-Zarr for the same reasons~\cite{moore2021omengff,moore2023omezarr}.
Large materials databases are also moving to the cloud.
For example, the Materials Project, which is built on \package{pymatgen}, is transitioning its entire data pipeline to cloud resources, including those of Amazon Web Services~\cite{jain2013commentary,horton2025accelerated}.

We keep all domains in a single repository, unlike the \package{HyperSpy} ecosystem, which has separated spectroscopy and file input and output into the \package{eXSpy} and \package{RosettaSciIO} packages~\cite{exspy,rosettasciio}.
Keeping all domains in one package means that data structures, physics models, representations, losses, and optimizers are written once and shared, so an improvement to a shared component reaches every method that uses it.
Combined methods such as ptychographic tomography can then import directly from several domains without version conflicts between packages, and we test every change to the core against all modules in one continuous integration run, at the cost of a larger package.

\subsection{\class{Dataset} and \class{Vector} Classes}
\label{sec:core}

\begin{figure}[tbp]
\centering
\includegraphics[width=6.4in]{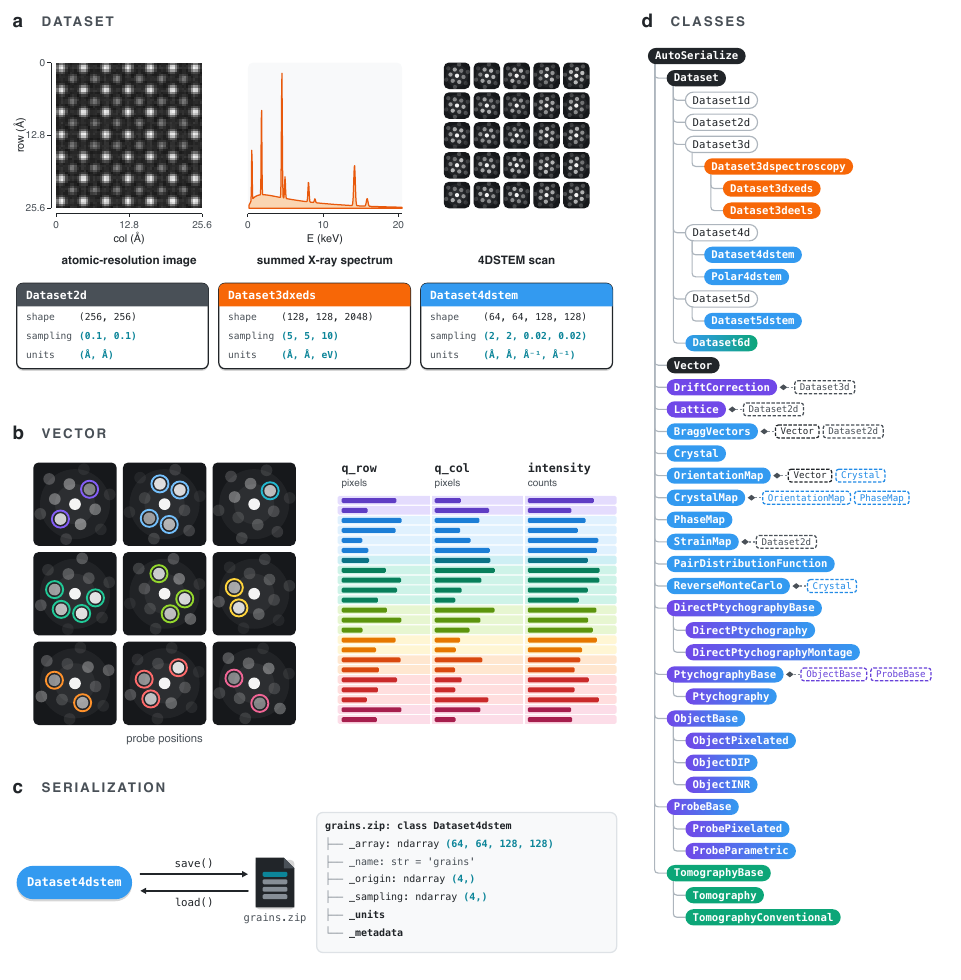}
\caption{Core data structures in \qem{}.
(a) An atomic-resolution image, an X-ray spectrum image shown by its summed spectrum, and a 4DSTEM scan, each with its calibrations.
(b) Bragg disks detected at nine probe positions and the \class{Vector} that stores them.
(c) Saving and loading a 4DSTEM dataset, and the file tree of the saved file.
(d) Selected classes that inherit \class{AutoSerialize}, colored by domain.
Tree lines show inheritance, and dashed chips show the classes stored as attributes.}
\label{fig:data_model}
\end{figure}

The \class{Dataset} class stores an N-dimensional array with the origin, sampling, and units of each axis, and slicing, cropping, binning, and Fourier resampling update these calibrations (Figure~\ref{fig:data_model}a).
We follow the axis convention of \package{RosettaSciIO}~\cite{rosettasciio}.
Derived data inherit these calibrations, so the mean diffraction pattern and virtual images of a 4DSTEM dataset are returned as \class{Dataset2d} objects in inverse \AA{}ngstr\"oms and \AA{}ngstr\"oms.
Subclasses of \class{Dataset} cover one- to four-dimensional data (\class{Dataset1d} to \class{Dataset4d}), 4DSTEM and polar-transformed 4DSTEM data (\class{Dataset4dstem} and \class{Polar4dstem}), and XEDS and EELS spectrum images (\class{Dataset3dxeds} and \class{Dataset3deels}), with five- and six-dimensional datasets in development (Figure~\ref{fig:data_model}d).
Each subclass adds methods specific to its measurement, such as virtual images and mean diffraction patterns for \class{Dataset4dstem}, but the same data can also be stored in the generic class of the same dimension, such as \class{Dataset4d}, which keeps the calibrations, slicing, and binning.
Each of these classes has a \code{show()} method that displays the data with calibrated axes.

\class{Dataset} and \class{Vector} inherit from the \class{AutoSerialize} base class, which saves and loads every attribute of an object (Section~\ref{sec:serialization}).
Analysis classes such as \class{BraggVectors}, \class{OrientationMap}, \class{StrainMap}, \class{Ptychography}, \class{Tomography}, and \class{DriftCorrection} inherit from the same base class, so we save and load them in the same way (Figure~\ref{fig:data_model}d).
Many of these classes also store other \qem{} classes as attributes; for example, \class{BraggVectors} stores the detected Bragg disks as a \class{Vector}, and \class{Ptychography} stores its object and probe models (Figure~\ref{fig:data_model}d).

Many analyses produce a variable number of entries at each probe position, such as detected Bragg disks.
The \class{Vector} class stores these ragged data on the fixed grid of probe positions, as the rows of one \package{PyTorch} tensor with named fields and units, together with the start and length of each position's rows (Figure~\ref{fig:data_model}b).
In Figure~\ref{fig:data_model}b, we color rows by probe position, and each bar shows the value of one field.
Indexing a \class{Vector} by probe position returns another \class{Vector}, and selecting fields returns a view that writes through to the shared buffer, so that operations such as calibrating the positions of all Bragg disks act on the whole buffer at once rather than looping over probe positions.
The buffer moves between the CPU and GPU like any other tensor.

We read vendor formats, including DigitalMicrograph, Velox EMD, Merlin MIB, and SER files, and standard formats such as Berkeley EMD, MRC, and TIFF, through the \package{RosettaSciIO} package with calibrations from the file metadata~\cite{rosettasciio}.
We can also read legacy \package{py4DSTEM} files and \package{abTEM} simulation outputs~\cite{savitzky2021py4dstem,madsen2021abtem}.

\subsection{Differentiable Forward Models}
\label{sec:forward}

\begin{figure}[tbp]
\centering
\includegraphics[width=6.4in]{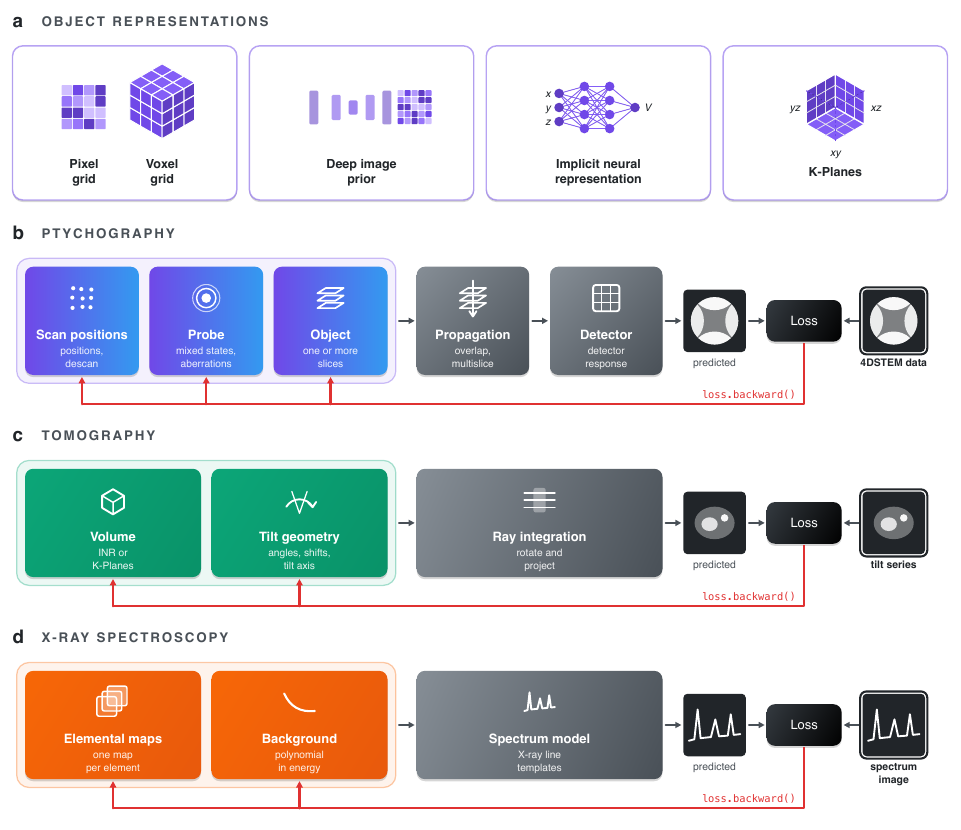}
\caption{Differentiable reconstruction examples in \qem{}.
(a) Object representations.
(b--d) Forward models for (b) ptychography, (c) tomography, and (d) X-ray energy-dispersive spectroscopy.
Colored: learnable components.
Gray: physics operators.
Black outline: measured data.
Red: gradients.}
\label{fig:autodiff}
\end{figure}

Each differentiable analysis in \qem{} is a forward model of learnable components and fixed physics operators (Figure~\ref{fig:autodiff}b--d).
For example, in ptychography, the scan positions, probe, and object are learnable, and multislice propagation and the detector response are physics operators (Figure~\ref{fig:autodiff}b).
We use a loss function to compare the predicted measurement with the measured data, and automatic differentiation sends the gradient of the loss back to every learnable component (red paths in Figure~\ref{fig:autodiff}b--d).
For tomography, the learnable volume is an INR or K-Planes decomposition projected along each tilt (Figure~\ref{fig:autodiff}c), and for XEDS, learnable elemental maps and a background polynomial in energy form a spectrum model (Figure~\ref{fig:autodiff}d).
By default, we compute gradients with autograd, so new losses, constraints, and components need no derived update rules.
For the most common ptychography loss, we also compute the gradient analytically, which matches autograd at lower cost, and preconditioning this gradient by the probe intensity reproduces the ePIE update~\cite{maiden2009improved,varnavides2023iterative}.
Each component holds its own \package{PyTorch} optimizer, learning-rate schedule, and constraints, so that, for example, probe positions can use a different learning rate than the object or be held fixed for the first iterations.
We add soft constraints such as total variation to the loss, and apply hard constraints such as positivity to the component output.
The loss library includes amplitude and intensity least squares, a Poisson likelihood, a smooth L1 loss, and a structural similarity loss~\cite{xie2023s3im}.

We use interchangeable representations, where any component can be a pixel or voxel grid, a deep image prior~\cite{ulyanov2018deep}, an INR~\cite{sitzmann2020implicit}, or a tensor decomposition~\cite{fridovichkeil2023kplanes} behind the same interface.
Switching between them changes one argument, which allows direct comparisons of regularization strategies on the same data~\cite{mccray2025deep,lim2025missing}.
The representations in Figure~\ref{fig:autodiff}a include a K-Planes decomposition built from three feature planes, and complex-valued layers let networks output wavefunctions and transmission functions directly.
The deep image prior is currently available for the ptychographic object, and tomography uses INR and K-Planes volumes.

Reconstructions run on CPUs, on CUDA and Apple silicon GPUs, and on several GPUs from a \package{Jupyter} notebook or under \code{torchrun}, with gradients averaged between workers~\cite{li2020pytorch}.
The same code runs on several nodes of high-performance computing (HPC) systems.
We have also implemented custom CUDA kernels to accelerate the most expensive operations in tomography, which also run under distributed training.

\subsection{Serialization}
\label{sec:serialization}

We can save any analysis object, from a dataset to a full reconstruction with its neural networks and optimizer states, to one \package{Zarr} zip file with \code{save()}.
We can then read these objects with \code{load()} (Figure~\ref{fig:data_model}c).
In reconstructions, we can store the path to the raw data and its preprocessing parameters instead of the raw data itself, which keeps saved files small.

\subsection{Visualization}
\label{sec:visualization}

\begin{figure}[tbp]
\centering
\includegraphics[width=6.4in]{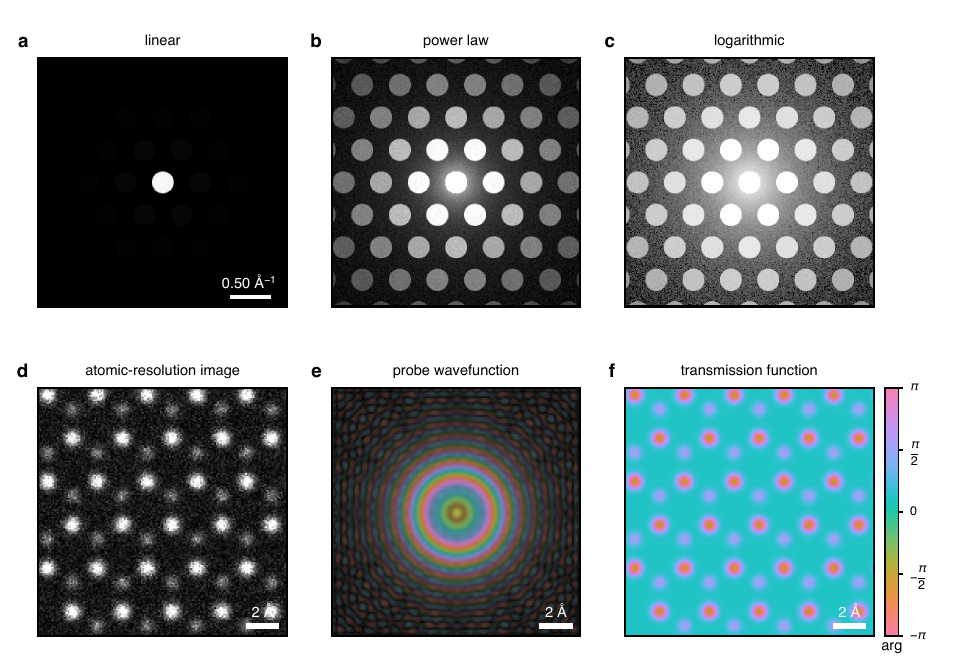}
\caption{Visualization of simulated data with \code{show\_2d()}.
(a--c) A nanobeam diffraction pattern with linear, power-law, and logarithmic normalizations.
(d) An atomic-resolution image.
(e) A STEM probe wavefunction.
(f) An object transmission function, with the phase colorbar for (e) and (f).}
\label{fig:visualization}
\end{figure}

Visualization has two levels: static \package{Matplotlib} figures for notebooks and publications~\cite{hunter2007matplotlib}, and interactive \package{anywidget} widgets for exploring large datasets~\cite{manz2024anywidget}.
For example, our function \code{show\_2d()} displays arrays, tensors, or datasets, alone or as a grid of panels, with per-panel normalization, colorbars, titles, and calibrated scale bars, and all six panels of Figure~\ref{fig:visualization} come from a single call.
We use intensity normalizations, adapted from \package{Astropy}~\cite{astropy2022}, to combine an interval (manual, centered, or quantile) with a stretch (linear, power-law, logarithmic, or inverse hyperbolic sine), as in Figure~\ref{fig:visualization}a--c.
We display complex-valued arrays by mapping amplitude to lightness and phase to hue in the perceptually uniform CIECAM02 color space, with an adjustable color saturation (Figure~\ref{fig:visualization}e,f).
We simulate the probe in Figure~\ref{fig:visualization}e at 300~kV with a 25~mrad aperture and 150~\AA{} defocus.
The phase of the transmission function in Figure~\ref{fig:visualization}f exceeds $\pi$ at the heavy atomic columns and wraps around the color wheel.

As an example of interactivity, the \class{Show4DSTEM} widget shows the diffraction pattern at a selected probe position and updates a virtual image as the user draws detectors in diffraction space, with data and GPU computation in Python and rendering in the browser.

\subsection{Building Blocks}
\label{sec:new}

\begin{figure}[tbp]
\centering
\includegraphics[width=6.4in]{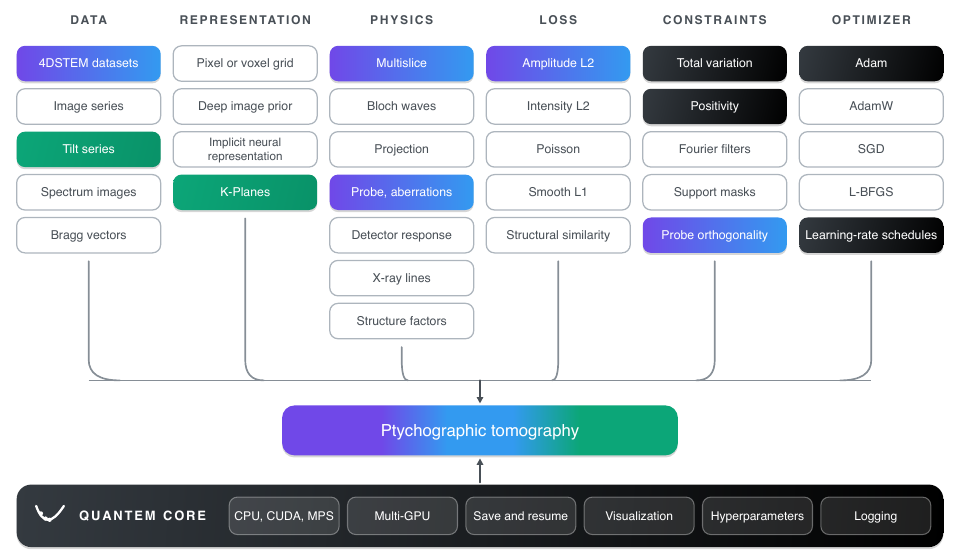}
\caption{Building blocks for new analysis methods in \qem{}.
Blocks used by ptychographic tomography are colored by source: ptychography (violet to blue), tomography (teal), and the \qem{} core (black).}
\label{fig:new_method}
\end{figure}

We assemble a new method from six types of building blocks, shown as the columns of Figure~\ref{fig:new_method}: data, representations, physics operators, loss functions, constraints, and optimizers.
We read crystal structures from CIF files or \package{ASE} \class{Atoms} objects~\cite{larsen2017atomic}, determine their space group symmetry with \package{spglib}~\cite{togo2024spglib}, and compute their reciprocal lattices and kinematical and dynamical structure factors in double precision, so that we can refine lattice parameters, strain, and orientation by gradient descent.
We compute electron scattering factors and projected potentials from a parameterization that obeys all physical constraints~\cite{lobato2014accurate}.
We model the loss of elastically scattered intensity to thermal diffuse scattering with absorptive form factors, which we compute numerically~\cite{weickenmeier1991computation}.

The differentiable probe model forms probes in real or Fourier space, including aberrations~\cite{kirkland2020advanced}, so that we can refine aberration coefficients in any reconstruction.
Calibration routines determine the diffraction pattern origin, the rotation between scan and diffraction coordinates, the diffraction pixel size, and the elliptical distortion of the diffraction plane.
We compute Fourier ring and shell correlations and the spectral signal-to-noise ratio from half datasets made by splitting the recorded electron counts, so that we evaluate every method with the same resolution metrics~\cite{varnavides2026beyond}.
The core also logs the progress of every method, and the diffractive imaging modules tune their hyperparameters by Bayesian optimization with \package{Optuna}~\cite{akiba2019optuna}.
Ptychographic tomography reuses the ptychography reconstruction with a voxel or K-Planes object that is rotated for each tilt before multislice propagation, using the blocks highlighted in Figure~\ref{fig:new_method}.

Other similar examples include full-pattern strain fitting, which models each diffraction pattern as a lattice of disk templates on a background, fits this differentiable model with the core optimizers, and measures strain from the fitted lattice vectors.
We have also implemented fused XEDS and HAADF tomography, which reuses the INR tomography reconstruction with a multichannel K-Planes object, a Poisson likelihood for the XEDS counts, and a chemical sparsity penalty.

\section{Current Modules}
\label{sec:modules}

\subsection{Crystal Orientation, Phase, and Strain}

We detect Bragg disks in 4DSTEM data by cross correlation with a vacuum probe template followed by subpixel refinement~\cite{pekin2017optimizing}.
We store the detected disks with calibrated positions and intensities, which we then use for indexing, lattice fitting, strain, and orientation mapping.
Automated crystal orientation mapping matches each diffraction pattern to a library of simulated patterns over all orientations~\cite{rauch2014automated}.
We previously accelerated this matching in \package{py4DSTEM} using a sparse correlation on the radial bands that contain reciprocal lattice points~\cite{ophus2022automated}, and in \qem{} the matching, refinement, and grain clustering all run as batched GPU operations.
We can also determine the phase fractions at each probe position from a nonnegative least-squares fit of several candidate crystal structures.

We measure strain from the positions of the Bragg disks relative to a reference lattice~\cite{pekin2017optimizing,ozdol2015strain}.
For thicker samples, we use Bloch wave calculations to refine the local thickness, tilt, and strain from the full intensity distribution of each pattern, extending the stacked Bloch wave approach~\cite{zeltmann2023uncovering}.
In multi-angle precession electron diffraction (MAPED), we record 4DSTEM data at several beam tilts, align the datasets in real and diffraction space, and combine them to reduce dynamical scattering effects~\cite{ribet2025maped}.

\subsection{Disordered and Semicrystalline Materials}

For amorphous and nanocrystalline materials, we compute the pair distribution function (PDF), the probability of finding two atoms at a given separation, from the radial diffraction intensity~\cite{ophus2023quantitative}.
The pair angle distribution function (PADF) extends this analysis to three-body correlations using angular correlations in the diffraction patterns~\cite{martin2017orientational}, and fluctuation electron microscopy measures medium-range order from the variance of the diffracted intensity~\cite{treacy1996variable}.
We also use reverse Monte Carlo fitting to refine a supercell model against diffuse scattering measured along several zone axes, which gives the Warren--Cowley short-range order parameters and atomic size effects.

\subsection{Phase Retrieval}

Diffractive imaging reconstructs the phase of the transmission function from 4DSTEM data, using either direct methods with closed-form solutions or iterative methods that solve an optimization problem~\cite{varnavides2026abcs}.
Direct methods include differential phase contrast and its integrated forms~\cite{close2015towards,lazic2016phase,yu2022realtime}, single sideband ptychography~\cite{rodenburg1993experimental,pennycook2015efficient}, optimum bright field imaging~\cite{ooe2021ultra}, matched illumination and detector interferometry~\cite{ophus2016efficient}, and parallax or tilt-corrected bright field imaging~\cite{varnavides2024tilt}.
In \qem{}, these direct methods share GPU kernels and fit the aberrations and scan rotation by cross correlation or least squares, with Bayesian hyperparameter search~\cite{akiba2019optuna}.
The shared framework includes upsampling, which relaxes the sampling requirements of direct ptychography~\cite{varnavides2026relaxing}, and the spectral signal-to-noise ratio, which provides a finite-dose metric for comparing phase retrieval methods~\cite{varnavides2026beyond}.
We can also reconstruct large fields of view by combining several scans, including ungridded and sparse scans, event-based detector data, and tilted samples with a varying defocus.

Iterative ptychography solves for the object and probe from overlapping diffraction patterns~\cite{faulkner2004movable,maiden2009improved,thibault2008high}, and can include mixed-state probes~\cite{thibault2013reconstructing}, multislice objects~\cite{maiden2012ptychographic}, and maximum-likelihood noise models~\cite{odstrcil2018iterative}.
Electron ptychography has reached deep sub-\AA{}ngstr\"om resolution in two-dimensional materials and thick crystals~\cite{jiang2018electron,chen2021electron}.
In \qem{}, iterative ptychography refines the scan positions, descan, background, probe, and single-slice or multislice object together, and each of these components can use a pixelated or neural representation~\cite{mccray2025deep}.
We also provide a simplified interface with default settings for common reconstructions.
Multislice ptychography in \package{py4DSTEM}, the predecessor of \qem{}, has resolved the three-dimensional structure of core-shell nanoparticles~\cite{ribet2024uncovering}.

\subsection{Atomic-Resolution Imaging}

Serial acquisition makes STEM images sensitive to sample drift and scan instabilities, which can be corrected by nonrigid registration of image series~\cite{jones2015smart} or by combining images recorded with orthogonal scan directions~\cite{ophus2016correcting}.
In \qem{}, drift correction aligns two or more images using translation, affine, and nonrigid models.

We fit lattice vectors to atomic-resolution images to define a reference for atom positions and strain.
Geometric phase analysis (GPA) measures displacement and strain fields from the phase of Bragg reflections in the Fourier transform of high-resolution images~\cite{hytch1998quantitative}.

\subsection{Three-Dimensional Structure}

Electron tomography reconstructs a three-dimensional volume from a tilt series of projections~\cite{midgley2001z,midgley2009electron}, and has reached atomic resolution~\cite{scott2012electron,miao2016atomic,pelz2023solving}.
\qem{} provides filtered back projection (FBP) and the simultaneous iterative reconstruction technique (SIRT)~\cite{gilbert1972iterative} on the GPU, as well as gradient-based reconstruction, which represents the volume as an INR or K-Planes decomposition and jointly refines the tilt angles and shifts of each projection~\cite{lim2025missing}.
The INR regularizes the reconstruction and fills in the missing wedge of unmeasured tilt angles~\cite{lim2025missing,kniesel2022clean}.
Gaussian splatting, a method adapted from novel view synthesis~\cite{kerbl2023gaussian}, represents a volume as a set of three-dimensional Gaussians whose positions, shapes, and amplitudes are optimized against the measured projections.
We can fit atomic positions to the reconstructed volumes, and analyze the resulting three-dimensional atomic models with radial distribution functions, nearest-neighbor statistics, polyhedral template matching, grain segmentation, and three-dimensional strain.

Ptychographic tomography reconstructs a three-dimensional potential from ptychographic datasets recorded at several tilt angles~\cite{dierolf2010ptychographic,pelz2023solving,varnavides2023iterative,lee2025phaset3m}, and in \qem{} we build it directly from the ptychography reconstruction (Section~\ref{sec:new}).
Scanning diffraction tomography reconstructs the local crystal orientation of each voxel from 4DSTEM datasets recorded at several tilts, using a multislice forward model.
Bend contour electron tomography (BCET) measures the strain and topography of thin crystalline films from the bend contours in a tilt series~\cite{bell2026bend}.

\subsection{Chemical Analysis}

We quantify XEDS spectrum images by identifying the X-ray lines, fitting backgrounds, denoising, generating elemental maps, and applying the Cliff--Lorimer method~\cite{cliff1975quantitative,watanabe2006quantitative}, or by fitting a model of the full spectrum image using gradient descent~\cite{lee2026plasmonic}.
We process EELS spectrum images by fitting power-law backgrounds, aligning the zero-loss peak, and estimating the relative thickness using the log-ratio method~\cite{egerton2011electron}.
XEDS and EELS tomography reconstruct three-dimensional elemental distributions from spectrum images recorded at several tilts~\cite{jarausch2009four,genc2013xeds}, and fusing the spectroscopic and HAADF signals improves the chemical resolution~\cite{schwartz2022imaging,schwartz2024imaging} (Section~\ref{sec:new}).

\subsection{Simulation}

The two standard methods for simulating elastic electron scattering are Bloch waves~\cite{bethe1928theorie} and the multislice method~\cite{cowley1957scattering,kirkland2020advanced}.
Using Bloch waves, \qem{} computes dynamical diffraction patterns, convergent-beam and large-angle convergent-beam electron diffraction patterns, and Kossel patterns with absorptive potentials.
We can also compute kinematical diffraction patterns with precession and convergent-beam excitation envelopes.
Multislice propagation is currently part of the forward models for multislice ptychography and scanning diffraction tomography, and tomographic projection operators provide the forward models for tomography.
We intend these simulations to be used as forward models for specific analyses, and recommend \package{abTEM} or \package{Prismatic} for general-purpose image and diffraction simulations~\cite{madsen2021abtem,rangeldacosta2021prismatic}.

\section{Discussion}
\label{sec:discussion}

\subsection{Comparison with Existing Software}
\label{sec:comparison}

\begin{table}[tbp]
\centering
\small
\caption{Comparison of \qem{} with other open-source Python packages for electron microscopy.}
\label{tab:comparison}
\setlength{\tabcolsep}{4pt}
\begin{tabularx}{\textwidth}{>{\raggedright\arraybackslash}p{2.5cm} >{\raggedright\arraybackslash}X >{\raggedright\arraybackslash}p{3.1cm} >{\raggedright\arraybackslash}p{1.3cm} >{\raggedright\arraybackslash}p{2.0cm} >{\raggedright\arraybackslash}p{1.7cm}}
\toprule
Package & Primary scope & Array backend & Autodiff & Native format & License \\
\midrule
\package{HyperSpy} & Multidimensional signals, model fitting~\cite{delapena2017electron,hyperspy} & \package{NumPy}, Dask & No & HDF5, \package{Zarr} & GPL-3.0 \\
\hspace{0.8em}\package{eXSpy} & EELS and XEDS~\cite{exspy} & \package{NumPy}, Dask & No & HDF5, \package{Zarr} & GPL-3.0 \\
\hspace{0.8em}\package{holospy} & Off-axis electron holography~\cite{holospy} & \package{NumPy}, Dask & No & HDF5, \package{Zarr} & GPL-3.0 \\
\hspace{0.8em}\package{RosettaSciIO} & File reading and writing~\cite{rosettasciio} & \package{NumPy}, Dask & No & HDF5, \package{Zarr} & GPL-3.0 \\
\hspace{0.8em}\package{pyxem} & 4DSTEM and diffraction~\cite{cautaerts2022free} & \package{NumPy}, Dask & No & HDF5, \package{Zarr} & GPL-3.0 \\
\package{py4DSTEM} & 4DSTEM~\cite{savitzky2021py4dstem} & \package{NumPy}, \package{CuPy} & No & HDF5 & GPL-3.0 \\
\package{LiberTEM} & Scalable and live processing~\cite{clausen2020libertem} & \package{NumPy}, \package{CuPy} & No & None & MIT \\
\package{pyTEMlib} & Spectroscopy, diffraction, and imaging in \package{pycroscopy}~\cite{somnath2019usid} & \package{NumPy}, Dask & No & HDF5 & MIT \\
\package{abTEM} & Simulation~\cite{madsen2021abtem} & \package{NumPy}, \package{CuPy}, Dask & No & \package{Zarr} & GPL-3.0 \\
\package{PtyRAD} & Electron ptychography~\cite{lee2025ptyrad} & \package{PyTorch} & Yes & HDF5 & LGPL-3.0 \\
\package{phaser} & Electron ptychography~\cite{gilgenbach2026phaser} & \package{NumPy}, \package{CuPy}, \package{JAX}, \package{PyTorch} & Yes & HDF5 & MPL-2.0 \\
\qem{} & Imaging, diffraction, tomography, spectroscopy, simulation~\cite{quantem} & \package{PyTorch} & Yes & \package{Zarr} & MIT \\
\bottomrule
\end{tabularx}
\end{table}

Table~\ref{tab:comparison} compares \qem{} with other open-source Python packages for electron microscopy.
The main difference is that \qem{} uses one differentiable backend for every modality, so that all reconstructions share the same optimizer framework, neural representations, and serialization, and we can assemble new methods from existing components.
\package{HyperSpy} and \package{RosettaSciIO} support a much wider range of signal types and file formats than \qem{}, and we use \package{RosettaSciIO} for file input.
\package{py4DSTEM} includes mature 4DSTEM analysis workflows that are not yet available in \qem{}, and \qem{} can read \package{py4DSTEM} files so that both packages can be used together.
\package{LiberTEM} provides distributed and live processing of very large datasets with user-defined functions, which \qem{} does not currently provide.
\package{PtyRAD}~\cite{lee2025ptyrad} and \package{phaser}~\cite{gilgenbach2026phaser} are dedicated electron ptychography packages that also use automatic differentiation, while \qem{} applies the same approach across imaging, diffraction, tomography, and spectroscopy.

\subsection{Contributing}
\label{sec:contributing}

We develop \qem{} on GitHub, where we merge contributions into a \code{dev} branch and publish releases to PyPI from the \code{main} branch.
Our contribution guidelines describe the development environment, code formatting, and pull request process, and are~\href{https://github.com/electronmicroscopy/quantem/blob/dev/CONTRIBUTING.md}{available online}.
Each new module includes tests and a tutorial notebook, and the tests run on Linux for Python 3.11 to 3.14 for every pull request, and on Windows and macOS for each release.
The citation file lists the software authors and the method papers for each module~\cite{smith2016software}.

\subsection{Development Status and Roadmap}
\label{sec:status}

\begin{table}[tbp]
\centering
\small
\caption{Development status of \qem{} components.}
\label{tab:status}
\begin{tabularx}{\textwidth}{>{\raggedright\arraybackslash}p{3.0cm} >{\raggedright\arraybackslash}X >{\raggedright\arraybackslash}p{2.4cm}}
\toprule
Layer or domain & Component & Status \\
\midrule
Core & \class{Dataset} for one- to four-dimensional data and 4DSTEM & Released \\
 & Conversion of storage from \package{NumPy} arrays to \package{PyTorch} tensors & In development \\
 & Ragged data (\class{Vector}, \package{PyTorch}-backed) & Released \\
 & Polar 4DSTEM and spectroscopy datasets & Released \\
 & Five- and six-dimensional datasets & In development \\
 & Serialization to \package{Zarr} & Released \\
 & Optimizers, constraints, losses & Released \\
 & Neural representations (CNN, SIREN, K-Planes) & Released \\
 & Multi-GPU ptychography & Released \\
 & \code{show\_2d()} and normalizations & Released \\
 & Interactive widgets (\class{Show4DSTEM}) & In development \\
\midrule
Physics & Crystallography, scattering factors, calibration & Released \\
 & Probe and aberration model & Released \\
\midrule
Imaging & Drift correction & Released \\
 & Lattice vector fitting & Released \\
\midrule
Diffraction & Bragg disk detection, orientation and phase mapping & Released \\
 & Bragg strain mapping, Bloch wave refinement & Released \\
 & Reverse Monte Carlo diffuse scattering & Released \\
 & Pair angle distribution function & In development \\
 & MAPED & Released \\
\midrule
Diffractive imaging & Iterative ptychography & Released \\
 & Direct ptychography & Released \\
 & Direct ptychography montage, event data & Released \\
 & Tilt-corrected dark field, PRISM ptychography & In development \\
\midrule
Tomography & SIRT, FBP, INR and K-Planes tomography & Released \\
 & Gradient-based voxel tomography & Planned \\
 & Gaussian splatting tomography & In development \\
 & Atom tracing and 3D atomic models & Released \\
\midrule
Spectroscopy & XEDS fitting and quantification & Released \\
 & EELS background, thickness & Released \\
\midrule
Simulation & Bloch wave, kinematical diffraction & Released \\
 & Standalone multislice & Planned \\
\midrule
Combined & Ptychographic tomography & In development \\
 & Scanning diffraction tomography & In development \\
 & XEDS/EELS tomography & In development \\
\bottomrule
\end{tabularx}
\end{table}

Table~\ref{tab:status} lists the development status of each component as of November 2026.
Released components are available in the \qem{} repository, components in development are in an open pull request or a feature branch, and we have not yet started the planned components.
The largest ongoing change is the conversion of all dataset storage from \package{NumPy} arrays to \package{PyTorch} tensors, so that every operation on a dataset can run on the GPU and be differentiable.
We are also adding the \qem{} version to every saved file, together with a safe loading mode that reads only arrays and tensors.

\section{Conclusions}
\label{sec:conclusions}

We have described \qem{}, an open-source Python package for quantitative electron microscopy built on \package{PyTorch} and \package{Zarr}, and its current modules across imaging, diffraction, tomography, spectroscopy, and simulation.
We implement analysis methods in \qem{} by writing a forward model and choosing representations, losses, constraints, and optimizers from shared components, so that every method gains GPU execution, serialization, and visualization from the core, and combined methods such as ptychographic tomography need no new gradient derivations or file conversions.
We encourage developers of new reconstruction methods to write them as differentiable forward models with shared data structures, so that they can be combined with other modalities and reused by other researchers.

\section*{Code and Data Availability}

\begin{sloppypar}
\qem{} is available at~\url{https://github.com/electronmicroscopy/quantem} under the MIT license and can be installed from PyPI with \code{pip install quantem}.
We archive each release on Zenodo~\cite{quantem}.
Tutorials are available at~\url{https://github.com/electronmicroscopy/quantem-tutorials}, and documentation at~\url{https://electronmicroscopy.github.io/quantem-docs/}.
Curated example datasets are available from the EM-Database at~\url{https://electronmicroscopy.github.io/emdatabase/}.
\end{sloppypar}

\section*{Acknowledgments}

CO and ARCM acknowledge funding support from the Accelerated Materials Design \& Discovery Program in the Toyota Research Institute.
GV acknowledges support from the Miller Institute for Basic Research in Science.
This material is based upon work supported by the U.S. Department of Energy, Office of Science, Office of Basic Energy Sciences, as part of the Genesis Mission National Science and Technology Challenges under Award Number DE-SCL0000271.
Work at the Molecular Foundry was supported by the Office of Science, Office of Basic Energy Sciences, of the U.S. Department of Energy under Contract No.~DE-AC02-05CH11231 under user proposal numbers MFP-9442, MFP-9619, and MFP-11179.

\bibliographystyle{\ifcompactinitials unsrtnat-initials\else unsrtnat\fi}
\bibliography{references}

\end{document}